\documentclass[twocolumn]{aastex631}

\usepackage{epstopdf}

\usepackage{newtxtext,newtxmath}
\usepackage{multirow}

\usepackage[T1]{fontenc}

\usepackage{graphicx}	
\usepackage{amsmath}	
\usepackage{amssymb}
\usepackage{epstopdf}
\usepackage{color}
\usepackage[caption = false]{subfig}

\shorttitle{Visual Binaries in Gaia eDR3 with 2MASS II}
\shortauthors{Medan, L\'{e}pine, Hartman \& Stassun}

\begin{document}
	
\title{Detecting New Visual Binaries in Gaia DR3 with Gaia and Two Micron All Sky Survey (2MASS) Photometry II. Speckle Observations of 16 Low-Separation Systems\footnote{Accepted 2024 April 03. Received 2024 April 01; in original form 2023 June 22}}

\author[0000-0003-3410-5794]{Ilija Medan}
\affiliation{Department of Physics and Astronomy, Georgia State University, Atlanta, GA 30302, USA}
\affiliation{Department of Physics and Astronomy, Vanderbilt University, Nashville, TN 37235, USA}

\author[0000-0002-2437-2947]{S\'{e}bastien L\'{e}pine}
\affiliation{Department of Physics and Astronomy, Georgia State University, Atlanta, GA 30302, USA}

\author[0000-0003-4236-6927]{Zachary Hartman}
\affiliation{Gemini Observatory/NSF’s NOIRLab, 670 A’ohoku Place, Hilo, HI 96720, USA}

\author[0000-0002-3481-9052]{Keivan G. Stassun}
\affiliation{Department of Physics and Astronomy, Vanderbilt University, Nashville, TN 37235, USA}


\begin{abstract}
Here we present speckle observations of 16 low-separation ($s<30$ AU) high probability candidate binaries from the catalog by Medan et al., where secondaries typically lack astrometric solutions in Gaia. From these speckle observations, we find a second component is always detected within the field of view. To determine if the detection is consistent with a physical companion or a chance alignment with a background source, we utilize a statistic from Tokovinin \& Kiyaeva that compares the apparent motion of the systems to the expected orbital motion ($\mu^\prime$). Using simulated binary orbits, we construct likelihood distributions of $\mu^\prime$ assuming various total errors on the measurements. With the hypothesis that the system is a true binary, we show that large measurement errors can result in $\mu^\prime$ values higher than expected for bound systems. Using simulated chance alignments, we also create similar likelihoods to test this alternative hypothesis. By combining likelihoods of both true binaries and chance alignments, we find that 15 of the 16 candidates are physical systems regardless of the level of measurement error. Our findings also accommodate all 16 as physical systems if the average, relative measurement error on the binary separations and position angles is $\sim4.3\%$, which is consistent with our knowledge of the Gaia and Gemini speckle pipelines. Importantly, beyond assessing the likelihood of a true binary vs. chance alignment, this quantitative assessment of the true average measurement error will allow more robust error estimates of mass determinations from short separation binaries with Gaia and/or Gemini speckle data.
\end{abstract}

\section{Introduction}

Large astrometric surveys have greatly expanded our knowledge of nearby, resolved binary systems. This is especially true for Gaia DR2/DR3 \citep{gaiadr2, gaiadr3}. Primarily, accurate parallaxes and proper motions allow for wide binaries to be detected in Gaia with a higher degree of certainty. This has led to numerous catalogs of wide binaries from Gaia totaling over 1 million systems \citep{elbadry2018, superwide, elbadry_2021}. These catalogs have worked well for binaries with separations greater than a few arcseconds. For smaller separations, such determinations are more difficult as the astrometry for the fainter secondary often has larger uncertainties or a solution cannot yet be found for the fainter component and is listed with no parallax or proper motion data.

The fact that Gaia DR3 lists pairs of sources with small angular separations ($<0.4$") in excess of the expected distribution from random field star alignments \citep{gaiaedr3_catalog_valid}, suggests that a large number of close separation (both on sky and physically) binary systems are missing from the current collections. To identify these barely resolved systems and distinguish them from chance alignments of unrelated sources, we have developed a method to assemble a catalog of 68,725 likely binaries within 200 pc that does not require the secondary component to have a measured proper motion or parallax \citep[][hereafter ``Paper I"]{medan2023}. Within this catalog, we find 590 previously unidentified binaries out of 696 candidate pairs with projected physical separations $< 30$ AU. For $s<10$ AU we find that 4 out of 15 candidate binaries have not yet been observed with high-resolution imaging. Systems with $s<30$ AU are of high interest for fundamental stellar astrophysics as their orbits can be monitored over a realistic time frame gravitational masses of the individual stars can be determined.

Speckle observations have been used over the last couple of decades to constrain the orbits and physical properties of binaries stars \citep[e.g.,][]{horch2008, tokovinin2010, Vrijmoet2022} as it allows for pairs to be resolved at the $\sim 10^1$ mas level. Our systems are at $\sim 10^2$  mas level of separation, but this will also allow for the detection of components with greater magnitude differences. Most of these past studies have also focused on observing brighter stars, where smaller telescopes can be used to detect the components of the system. In this study we will be targeting fainter candidate primaries ($G \gtrapprox 13$) as these have not been systematically targeted by past programs, meaning we will be using the larger Gemini $8.1$ m North and South telescopes to carry out our observations.

In this study, we present observations for sixteen of these close physical separation systems that have not previously been identified. The goals here are not only to confirm the binary status of these systems so they can continually be monitored to map their orbit, but also to obtain an estimate of the contamination rate of chance alignments in our catalog Paper I. This is an important statistic to asses whether our catalog may be useful in future studies. Additionally, we will show that these data can be used to statistically define the average level of error on the positions from Gaia and/or the Gemini speckle instruments. This result is crucial for future orbital solutions of visual binaries that rely on data similar to that presented here.

The structure of this paper is as follows. In Section \ref{sec:data} we discuss how the 16 candidate binaries were selected and describe the results from the speckle observations. In Sections \ref{sec:orb_mot}$-$\ref{sec:prop_back} we describe the methods we use to determine if the motion of the secondary between the epoch of Gaia and the speckle observations is due to orbital motion, or if it is more consistent with chance alignment with a background star given an assumed level of error on the measurements. In Sections \ref{sec:trips} and \ref{sec:multi} we discuss the detection of a third component in two of the observations, and further examine another system for which multiple epochs of speckle observations have been obtained; we discuss the implications from both results. Finally in Section \ref{sec:discuss} we calculate the rate of binaries detected in this sample and how this compares to the estimated contamination rate from Paper I. Here we also discuss how the possible level of error effects these results and what issues this may cause in future studies.

\section{Data}\label{sec:data}

\subsection{Target Selection}\label{sec:targ_selc}

For this study, we are attempting to confirm select candidate binaries from Paper I, in which we identified 68,725 likely candidate binaries within 200 pc from the Gaia DR3 catalog. These binaries were selected without the use of Gaia astrometry, so follow-up observations are needed to (1) confirm binarity through relative proper motion measurement, and (2) additionally validate the methodology from our Paper I study. The full details of the method can be found in Paper I, but we provide a summary of the method here for the reader.

In Gaia DR3, there is an excess of objects at small separations compared to past data releases \citep{gaiaedr3_catalog_valid}. It is difficult to determine if these close neighbors are spurious entries in the catalog (e.g. duplicates), or whether they are true detections of resolved sources, especially given that 74\% of neighbors at separations $<0.4$" only have a 2-parameter (position only) solution \citep{gaiaedr3_catalog_valid}. Because of this, we developed a method to identify binaries without the use of Gaia parallax and proper motion values. To accomplish this, we compared the multidimensional distribution consisting of the $G$ magnitude of the primary ($G$), sine of the Galactic latitude ($sin(b)$), Galactic longitude ($l$), $|\Delta G|$, $G-J$ excess ($(G - J) -  (G - J)_0$), angular separation ($\theta$) and the \texttt{ipd\_frac\_multi\_peak} value from \textit{Gaia}. These distributions were constructed for close pairs within 200 pc ($<2.5$") and for likely chance alignments, where chance alignments are defined as pairs with either $|\pi_1 - \pi_2| / \left( \sqrt{\sigma_{\pi_1}^2 + \sigma_{\pi_2}^2} \right) > 6$ or the separation between primary and secondary is consistent with the field star distribution according to Poisson statistics. We then use the densities from the two distributions, $N_{cand}(\overrightarrow{x})$ for all 200 pc candidates and $N_{chance}(\overrightarrow{x})$ for likely chance alignments, to calculate a value analogous to a likelihood of a pair of stars being a chance alignment, which in Paper I we referred to as a ``contamination factor":
\begin{equation}\label{eq:likelihood}
	L = \frac{N_{chance}(\overrightarrow{x})}{N_{cand}(\overrightarrow{x})}
\end{equation}
The above is not strictly a probability, as not all values fall between 0 and 1. To calibrate our contamination factor, we compare the sky distribution of likely binaries with $L < L_{cut}$ to the expected sky distribution for 200 pc stars. Over- or under-densities compared to the expected distribution signify contamination or completeness issues at a given cut. This allowed us to select an $L_{cut}$ that struck a balance between completeness and contamination. With an optimal selection of $L < 0.00193$, we recover true binaries in our ``clean sample" with a completion rate of $\sim 64\%$ and a contamination rate of $\sim 0.4\%$.

\begin{figure*}[!t]
	\centering
	\plotone{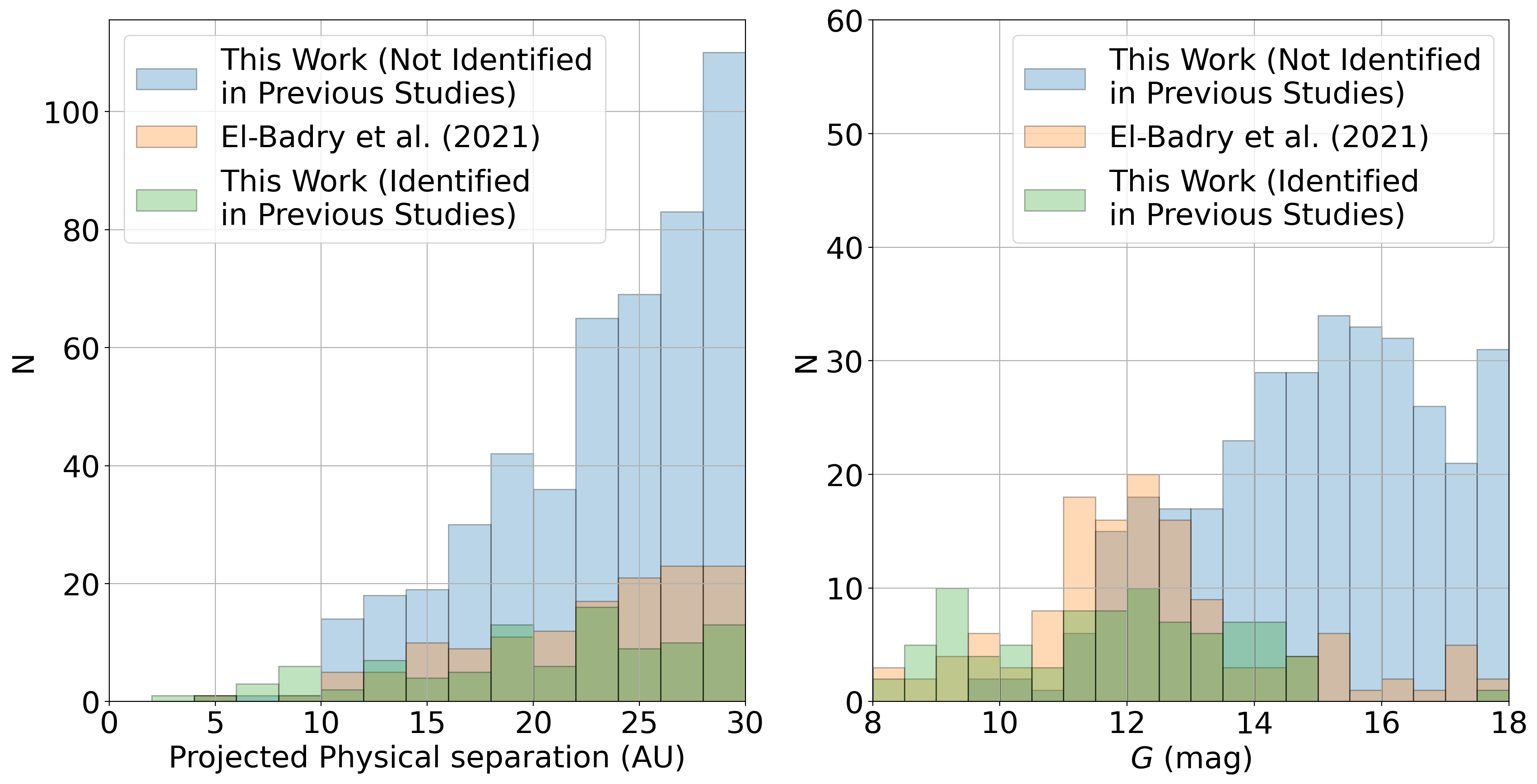}
	\caption{Distribution of projected physical separations (left panel) and Gaia $G$ magnitude (right panel) for candidate binary stars resolved by Gaia eDR3. The blue histograms are for candidate binaries identified with our method with $s<30$ AU, and that have not been identified as binaries in previous studies. The orange histograms are for binaries identified in \cite{elbadry_2021} with $d<200$ pc and $s<30$ AU. The green histograms are for candidate binaries identified with our method but already listed in the Washington Double Star Catalog \citep{WDS} or observed with high resolution imaging in previous studies.}
	\label{fig:sep_G_dist}
\end{figure*}

For this study we focus on the closest physical separation binaries ($s<30$ AU) identified in Paper I, as these systems are the best for long-term astrometric monitoring to map out significant portions of their orbits in a reasonable time-span, from which Keplerian mass estimates can be derived. These systems are plotted in Figure \ref{fig:sep_G_dist}, where the blue histograms show the candidate binaries identified with our method with $s<30$ AU and that had not been identified as binaries in previous studies, the orange histograms are for binaries identified in \cite{elbadry_2021} with $d<200$ pc and $s<30$ AU, and the green histograms are for candidate binaries identified with our method but are already listed in the Washington Double Star Catalog \citep{WDS} or observed with high resolution imaging in previous studies. Here we see that our method has the potential to greatly increase the number of known nearby binaries with $10<s<30$ AU. Additionally, we see that our candidate binaries tend to be much fainter than those previously identified, requiring larger telescopes to confirm the systems through follow-up with ground-based, high angular resolution methods like AO or speckle. For this study we focus on the closest physical separation binaries that are near the fainter end of the systems that have previously observed, where the use of larger telescopes is most warranted. The 16 candidate binaries observed for this study are summarized in Table \ref{tab:gaia_data}, where we include the distance and photometric information for the primary, and the magnitude difference in Gaia between the primary and secondary.

\begin{table*}
	\centering
	\caption{Corresponding Gaia data for the 16 binaries targeted in this study. Below the distance and photometric data listed is for the primary, and the magnitude difference listed is between the primary and secondary as identified in \cite{medan2023}.}
	\label{tab:gaia_data}
	\begin{tabular}{llccccccc} 
			\hline
			Binary Number & Gaia DR3 \texttt{source\_id }& R.A. & Decl. & Distance & $G_1$ & $BP_1$ & $RP_1$ &$|\Delta G|$ \\
			& & [deg.] & [deg.] & [pc] & [mag] & [mag]& [mag]& [mag]  \\
			\hline
			1 & 3229551253220231808 & 70.90087 & -0.56731 & 102.040 & 14.212 & 15.369 & 12.429 & 0.273\\
			2 & 5492027049935313664 & 110.11636 & -52.30629 & 42.678 & 13.768 & 14.653 & 11.971 & 0.292\\
			3 & 6261708725800628864 & 236.48278 & -16.54947 & 26.633 & 13.326 & 14.600 & 12.001 & 1.492\\
			4 & 6411063808333667328 & 333.94860 & -58.94506 & 21.340 & 13.123 & 14.322 & 11.300 & 0.344\\
			5 & 6644840974200903808 & 289.27319 & -52.64863 & 23.814 & 13.815 & 15.302 & 12.127 & 0.774\\
			6 & 4953865551800209408 & 39.98967 & -36.42878 & 41.406 & 14.056 & 14.731 & 12.323 & 0.142\\
			7 & 2417948085206509952 & 2.50716 & -13.91104 & 51.927 & 13.829 & 14.553 & 11.972 & 0.030\\
			8 & 6613996340143809664 & 335.18743 & -30.87538 & 42.310 & 13.873 & 14.704 & 12.075 & 0.167\\
			9 & 4933729022833362176 & 19.96248 & -47.67816 & 59.070 & 14.565 & 15.351 & 12.742 & 0.004\\
			10 & 4927117320114483584 & 17.46732 & -53.00546 & 68.622 & 14.348 & 15.126 & 12.545 & 0.366\\
			11 & 6904650039925953792 & 308.38467 & -9.98559 & 39.312 & 13.399 & 14.204 & 11.481 & 0.014\\
			12 & 5531191997720282240 & 119.57536 & -44.73889 & 31.447 & 12.711 & 13.489 & 10.761 & 0.096\\
			13 & 4723513708252878080 & 46.08146 & -60.43128 & 35.510 & 13.886 & 14.907 & 12.026 & 0.097\\
			14 & 3053294041735602688 & 106.27457 & -5.53672 & 58.390 & 13.837 & 14.373 & 12.058 & 0.012\\
			15 & 2925481999752178688 & 100.54701 & -23.10836 & 46.723 & 13.129 & 13.827 & 11.281 & 0.025\\
			16 & 5390947154990629248 & 161.11498 & -43.54761 & 33.108 & 12.921 & 13.800 & 11.084 & 0.196\\
			\hline
	\end{tabular}
\end{table*}

\subsection{Speckle Imaging}\label{sec:seckle}

We observed a total of 16 candidate binaries using the ‘Alopeke and Zorro speckle cameras \citep{scott2018, scott2021} on the Gemini $8.1$ m North and South telescopes, respectively, from February 2022 to January 2023. These observations were part of the Gemini observing programs GN-2022A-Q-313, GS-2022A-Q-316 and GS-2022B-Q-313, and were taken following the NASA speckle team's procedure. In short, observations are taken with the ‘Alopeke and Zorro cameras, which consist of a 256x256 pixel array. Observations are taken simultaneously in two bands, a ``blue" band centered on 562 nm and a ``red" band centered on 832 nm. An observation consists of multiple sets of 1000 images of 60 milliseconds, with each set equaling one minute of total integration time.  The number of sets depends on the magnitude of the object and the weather conditions at the time of observation.  The pixel scale of each band is $\sim 0.0099"$ and $\sim0.0109"$, respectively, though the pixel scale is re-calibrated each observing run, as we will discuss in more detail below.

\begin{figure*}[!t]
	\centering
	\includegraphics[width=\textwidth]{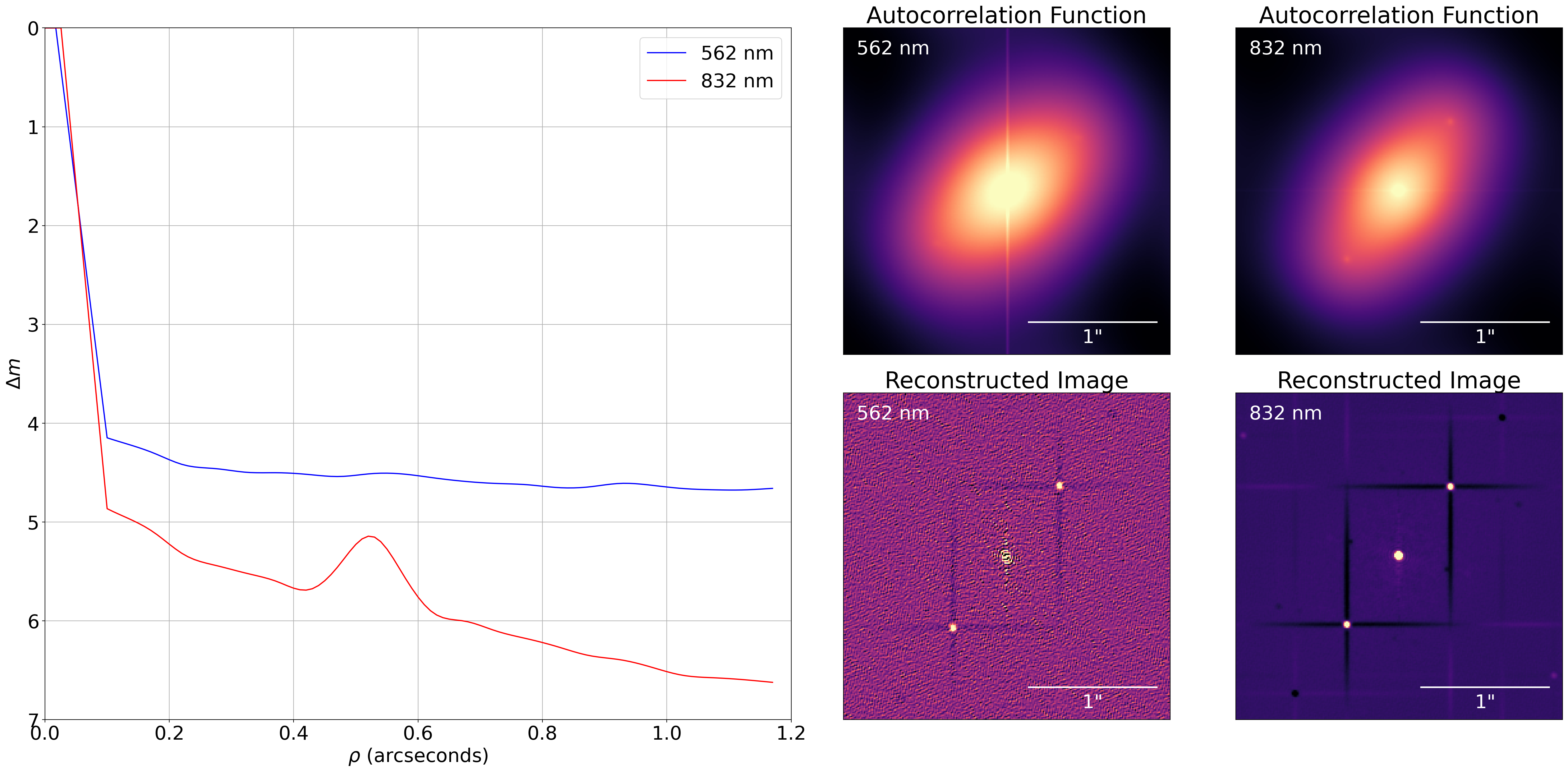}
	\caption{Example of the data products produced by the Gemini speckle pipeline for Binary 2 in Table \ref{tab:speckle_results}. The red and blue lines in the far left panel show the contrast limit for the 832 nm and 562 nm bands, respectively, which is a result of using the background flux levels to determine the faintest companions one can reliably detect as a function of separation. In the four panels on the right, we show the autocorrelation functions (top row) and reconstructed images (bottom row) for the binary in each bandpass, where the bandpass is indicated in the top left corner of each panel.}
	\label{fig:ex_speckle}
\end{figure*}

The data from these observations were reduced by the Gemini team using the pipeline originally created by \cite{horch2011} and later modified by \cite{howell2011}. The full details of the pipeline can be found in \cite{horch2011} and \cite{howell2011}, but a brief description is provided here. First, the pipeline calculates the autocorrelation for each frame in the observation. For images consisting of pairs of speckles (one for each star), the autocorrelation should result in one central peak where the frame is aligned with itself, and two additional lower level peaks where one speckle pattern is correlated with the other. To build signal, the autocorrelations for all frames of a binary are summed. An example of an autocorrelation for one of our candidate binaries is shown in Figure \ref{fig:ex_speckle}. Next the pipeline calculates a power spectrum by taking the Fourier transform of the summed autocorrelations. The power spectrum of the candidate binary is divided by the power spectrum of a standard point source to acquire an image of the fringe patterns of the science target. This fringe pattern is then used to reconstruct the image of the candidate binary, where an example of a reconstructed image is shown in Figure \ref{fig:ex_speckle}. From this reconstructed image, the apparent separation and position angle of the secondary is used to get an initial guess of the solution. This initial guess is used to find the best fit model of the fringe pattern, using the method from \cite{horch1996}. The pipeline team then compares the best fitted fringe pattern to the observed one and examines the resulting reconstructed images. This allows the speckle team to intervene and either flag poor fits or attempt to interactively improve the modeled fringe pattern. The final results are then calculated from this best fit model fringe pattern, where the separation ($\rho$) and position angle ($\theta$) are determined from the spacing and orientation of the fringe pattern, respectively. The magnitude difference in each band is also determined from the amplitude of the model fringe pattern.

To determine if the detection found from the above pipeline analysis is significant, the contrast limit for each observation is determined. To find this, the distribution of local minima and maxima in the background of the reconstructed images are calculated. Here, the standard deviation of these minima and maxima from the mean background are found within a series of annuli from the central star. The detection limit is then set to be $5\sigma$ brighter than this mean background within each annulus. An example of such a contrast curve for one of our candidate binaries is shown in Figure \ref{fig:ex_speckle}. From the example of these pipeline results, it is clear that for $\rho > 0.2"$, secondaries can be detected with magnitude differences $>4$ mag. This is much greater than what we expect for all of the candidate binaries in this work. We do note that in most reconstructed images resulting from the pipeline, we notice an artifact of a cross pattern through the secondary, similar to those shown in Figure \ref{fig:ex_speckle}. This is a known artifact that can sometimes occur in low signal-to-noise observations \citep{howell2012}, which our observations fall in the category of due to being fairly faint. As the solutions presented here are wholly based on the fringe pattern in Fourier space, these artifacts have no effect on the final solution. This demonstrates how these reconstructed images should be used for visual purposes only.

One crucial aspect of speckle imagining is the astrometric calibration, which is achieved by measuring the pixel scale and orientation for each observing run using a set of calibrator binaries. These calibrator binaries have measured orbital parameters, such that the separation and position angle of the system is known at the time of observation. During an `Alopeke or Zorro run, several calibrator binaries are typically observed using the procedure described above. Reducing this data through the standard speckle pipeline, the resulting separation and position angle for both cameras are then saved in pixel units. This solution in pixel units is compared to the true separation and position in physical units, where the comparison of the separation provides the pixel scale for each camera and the position angle the orientation of the array. This procedure is repeated multiple times over the observing run with multiple calibrator binaries to get the average pixel scale and orientation.

\begin{figure*}[!t]
	\centering
	\includegraphics[width=0.95\textwidth]{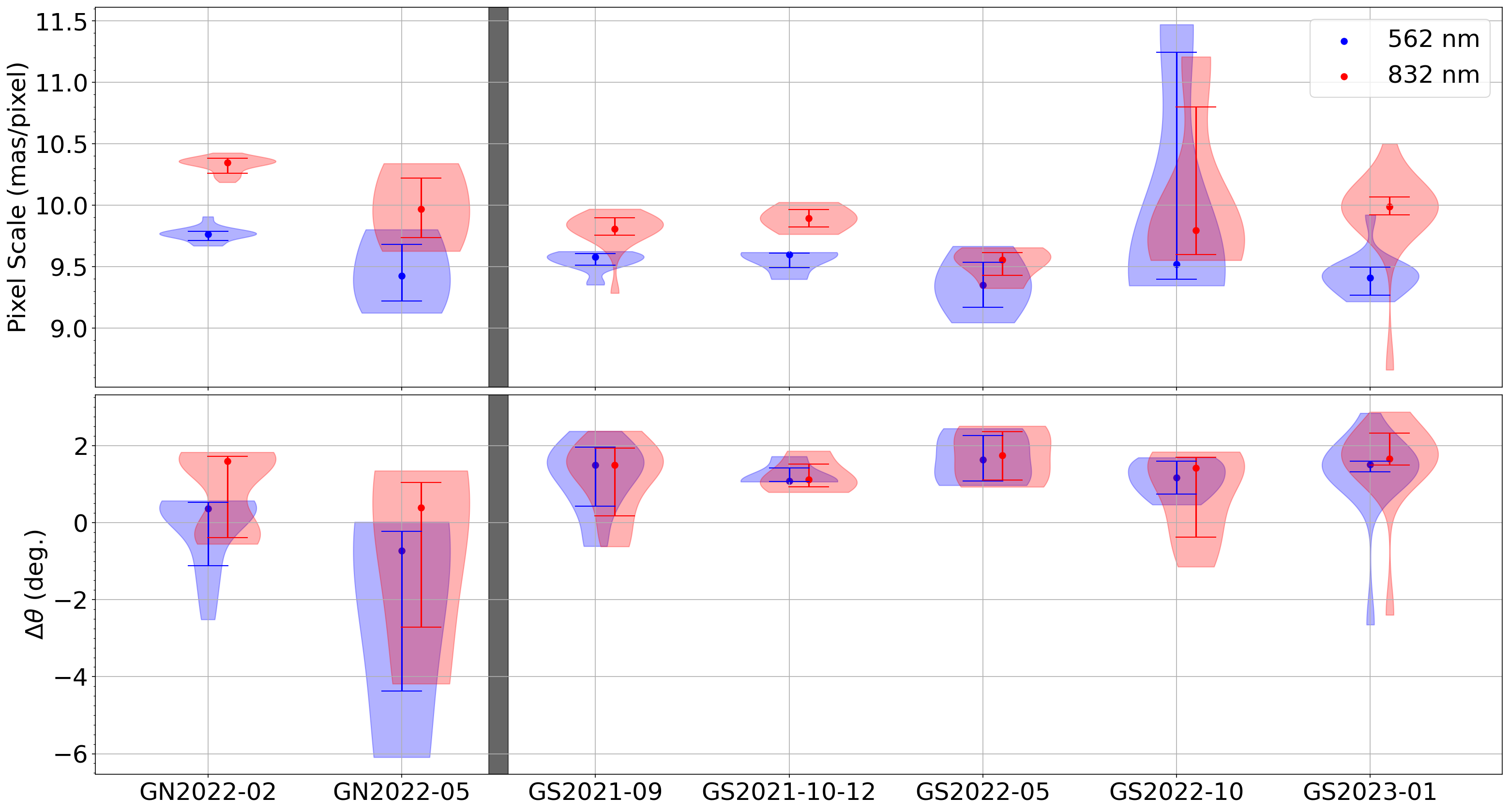}
	\caption{Violin plot of the pixel scales (top row) and offset in the position angle from the true value for the calibration binaries (bottom row). In each plot, the blue violin plots are for the 562 nm camera and the red for the 832 nm camera. The data points and error bars show the 50th and 16th/84th percentiles, respectively. These percentiles correspond to the values shown in Table \ref{tab:astro_calib}. The shaded regions in the plots show the distribution of the data, as estimated by a Gaussian Kernel Density Estimator. In both rows, the solid black bar separates the astrometric calibrations for ‘Alopeke (GN in the x-axis labels) and Zorro (GS in the x-axis labels). In each label, the year and month of the observing run is shown.}
	\label{fig:asrtro_calib}
\end{figure*}

Table \ref{tab:astro_calib} and Figure \ref{fig:asrtro_calib} show a summary of these results for each observing run over the time span of the observations in the current study. Here, we show the median and 16th/84th percentile of the pixel scale for each camera, and the median and 16th/84th percentile of the difference between the true and measured position angle of the calibrator binaries after the correct rotation had been applied to the array. Additionally, the ``violin plots" in Figure \ref{fig:asrtro_calib} show the distribution of the data, as estimated by a Gaussian Kernel Density Estimator. Here some distributions may seem truncated or asymmetric due to the small number of observations for a given observing run (see Table \ref{tab:astro_calib}). Between each observing run, we expect small changes in the pixel scale due to e.g., adjustment of the science fold mirror on the telescope \citep{scott2021}. We do see that within an observing run, that the pixel scale does have some level of variation. For most observing runs, this variation is $< 1$ mas though. Larger variations can be due to things like poor weather conditions during a specific observing run, and could result in larger variances like what is seen for GS2022-10. When considering the difference in the corrected position angles, we find that the precision is at the level of up to a few degrees, but typically $\sim1^\circ$. Additionally from these calibrations, we find that there is some difference in the position angle between the red and blue cameras, typically at the level of $<1^\circ$ for Gemini North and $<0.2^\circ$ for Gemini South.

\begin{table*}
	\centering
	\caption{The resulting pixels scales and offset in the position angle from the true value for the calibration binaries during each observing run. Here observing runs are labeled by observatory, where ‘Alopeke is labeled with GN and Zorro with GS, and the year and month of the observing run. The last two columns of the table show the number of unique calibration binaries observed during the run ($N_{binaries}$) and the total number of observations of all of the binaries ($N_{obs}$).}
	\label{tab:astro_calib}
	\begin{tabular}{lcccccc}
		\hline
		Observing Run & Pixel Scale (562 nm) & Pixel Scale (832 nm) & $\Delta \theta_B$ & $\Delta \theta_R$ & $N_{binaries}$ & $N_{obs}$ \\
		                       & [mas/pix.]                     & [mas/pix.]                      & [deg.]                  & [deg.] & &  \\
		\hline
		GN2022-02 & $9.7651^{+0.0228}_{-0.0534}$ & $10.3456^{+0.0348}_{-0.0843}$ & $0.3660^{+0.1580}_{-1.4849}$ & $1.5921^{+0.1338}_{-1.9792}$ & 5 & 18\\
		GN2022-05 & $9.4252^{+0.2561}_{-0.2050}$ & $9.9683^{+0.2524}_{-0.2324}$ & $-0.7302^{+0.5080}_{-3.6449}$ & $0.3890^{+0.6516}_{-3.1085}$ & 2 & 3\\
		\hline
		GS2021-09 & $9.5783^{+0.0263}_{-0.0665}$ & $9.8085^{+0.0898}_{-0.0520}$ & $1.5007^{+0.4536}_{-1.5285}$ & $1.4960^{+0.4371}_{-1.3208}$ & 8 & 17\\
		GS2021-10-12 & $9.5990^{+0.0119}_{-0.1073}$ & $9.8940^{+0.0711}_{-0.0712}$ & $1.0773^{+0.3373}_{-0.0117}$ & $1.1198^{+0.4056}_{-0.1925}$ & 3 & 4\\
		GS2022-05 & $9.3485^{+0.1872}_{-0.1798}$ & $9.5570^{+0.0569}_{-0.1268}$ & $1.6295^{+0.6262}_{-0.5455}$ & $1.7494^{+0.6110}_{-0.6431}$ & 5 & 5\\
		GS2022-10 & $9.5180^{+1.7279}_{-0.1218}$ & $9.7952^{+1.0020}_{-0.1985}$ & $1.1625^{+0.4267}_{-0.4266}$ & $1.4151^{+0.2766}_{-1.7942}$ & 8 & 8\\
		GS2023-01 & $9.4113^{+0.0838}_{-0.1455}$ & $9.9871^{+0.0811}_{-0.0641}$ & $1.5124^{+0.0843}_{-0.1924}$ & $1.6618^{+0.6676}_{-0.1615}$ &6 & 16\\
		\hline
	\end{tabular}
\end{table*}

For this study, the following observing runs were used to reduce our data. For Gemini North data taken with the 'Alopeke speckle camera, all observations from February to March 2022 were reduced with the GN2022-02 data. This includes Binaries $1-3$. The rest of the binaries were observed at Gemini South with the Zorro speckle camera. Here the data taken from March 2022 were calibrated with the observations from GS2021-09 and GS2021-10-12, and the data from May 2022, October 2022  and January 2023 were calibrated with GS2022-05. We do note that the Gemini speckle team chose in many instances to not calibrate the pixel scale and array orientation with the data from the same observing run. This is because for some observing runs, less calibration binaries were observed. This decision is made as, generally, the primary goal of most programs using the speckle instrument is the detection of objects and not the high accuracy of their position. Additionally, as the instrument is permanently mounted on the telescope, there is little change in the pixel scale and orientation over time. We find this to be generally true here (see Figure \ref{fig:asrtro_calib}), again at the $< 1$ mas level in pixel scale and $<1^\circ$ level in array orientation. We do note here that such errors in the calibration are multiplicative in regards to the errors on the final separation and position angle. The implications of this will be discussed in more detail in Section \ref{sec:discuss}.

With the above in mind, the speckle pipeline results for our candidate binaries are summarized in Table \ref{tab:speckle_results} for both the blue band (562 nm) and red band (832 nm) observations. For reference, the separation and position angles at the Gaia epoch are also included. As a note, the pipeline also reports flags for where the quadrant of the detection is ambiguous or there was a difficult fit in the reduction. For the vast majority of the detections, the quadrant of the source is found to be ambiguous and could be off by $180^\circ$.

\begin{table*}
	\footnotesize
	\centering
	\caption{Speckle imaging results for the 16 candidate binaries, indicated by their Binary Number from Table \ref{tab:gaia_data}. Both the detected separation ($\rho$) and position angle from the brighter target to the fainter companion ($\theta$, North through East) and the magnitude difference between the detected sources in the blue band (562 nm) and red band (832 nm) are shown. If no result is shown, than no detection was made in this band. For reference, included in the 2nd through 4th columns are the separation, position angle and magnitude difference of the binary according the the Gaia catalog. The last column of the table shows any flags from the reduction.}
	\label{tab:speckle_results}
	\begin{tabular}{lccccccccccc} 
		\hline
		Binary Number & $\rho_G$ & $\theta_G$ & $\Delta G$ & Obs. Date & $\rho_B$ & $\theta_B$ & $\Delta B$ & $\rho_R$ & $\theta_R$ & $\Delta R$ & Flags \\
		& [arcseconds] & [deg.] & [mag] & [YYYY/MM/DD] & [arcseconds] & [deg.] & [mag] & [arcseconds] & [deg.] & [mag] & \\
		\hline
		1 & 0.279 & 316.079 & 0.273 & 2022/02/11 & 0.439 & 320.669 & 1.400 & 0.443 & 321.415 & 0.770 & \nodata\\
		2 & 0.627 & 152.228 & 0.292 & 2022/03/16 & 0.662 & 324.239 & 1.310 & 0.662 & 324.267 & 0.430 & \nodata\\
		3 & 1.067 & 246.272 & 1.492 & 2022/03/17 & 1.122 & 244.466 & 2.310 & 1.119 & 244.099 & 1.390 & \nodata\\
		4 & 0.753 & 151.189 & 0.344 & 2022/05/18 & 0.623 & 337.966 & 1.920 & 0.621 & 338.142 & 0.480 & a\\
		4b$^\dagger$ & \nodata & \nodata & \nodata & 2022/05/18 & \nodata & \nodata & \nodata & 1.034 & 226.040 & 6.820 & a,b\\
		5 & 0.814 & 125.106 & 0.774 & 2022/05/23 & 0.888 & 126.288 & 1.840 & 0.887 & 126.768 & 0.760 & \nodata\\
		6 & 0.308 & 192.160 & 0.142 & 2022/10/07 & 0.293 & 347.031 & 1.170 & 0.288 & 346.428 & 0.500 & a\\
		7 & 0.277 & 130.912 & 0.030 & 2022/10/07 & 0.288 & 335.778 & 1.500 & 0.281 & 334.903 & 0.380 & a\\
		8 & 0.248 & 294.830 & 0.167 & 2022/10/08 & \nodata & \nodata & \nodata & 0.284 & 99.369 & 0.750 & a\\
		9 & 0.238 & 224.183 & 0.004 & 2022/10/08 & \nodata & \nodata & \nodata & 0.491 & 220.735 & 1.090 & a\\
		4 & 0.753 & 151.189 & 0.344 & 2022/10/08 & \nodata & \nodata & \nodata & 0.609 & 338.109 & 1.250 & a\\
		10 & 0.251 & 187.824 & 0.366 & 2022/10/09 & 0.464 & 188.418 & 2.510 & 0.475 & 187.630 & 0.740 & a\\
		11 & 0.405 & 138.193 & 0.014 & 2022/10/09 & 0.452 & 152.782 & 1.080 & 0.442 & 152.731 & 0.280 & a\\
		12 & 0.348 & 123.636 & 0.096 & 2022/11/08 & 0.327 & 351.146 & 0.590 & 0.321 & 350.826 & 0.220 & a\\
		13 & 0.500 & 197.058 & 0.097 & 2022/11/08 & 0.595 & 198.878 & 1.030 & 0.589 & 198.777 & 0.260 & a\\
		14 & 0.238 & 227.535 & 0.012 & 2023/01/08 & 0.316 & 38.634 & 1.480 & 0.299 & 39.527 & 0.380 & a\\
		15 & 0.364 & 55.078 & 0.025 & 2023/01/08 & 0.234 & 204.083 & 0.560 & 0.224 & 204.320 & 0.200 & a\\
		16 & 0.510 & 334.019 & 0.196 & 2023/01/09 & 0.596 & 336.266 & 0.620 & 0.569 & 336.217 & 0.290 & a,b\\
		16b$^\dagger$ & \nodata & \nodata & \nodata & 2023/01/09 & \nodata & \nodata & \nodata & 1.151 & 41.948 & 9.630 & a,b\\
		\hline
	\end{tabular}
	\\
	$^\dagger$Speckle detection of an apparent third component in the field that was not detected in the original Gaia-2MASS binary analysis.
	\\
	$^\text{a}$The quadrant of the position angle is ambiguous, meaning that it could be off by 180 degrees. Here, when possible, the quadrant deemed most probable is selected.
	\\
	$^\text{b}$A difficult fit or one with a goodness-of-fit that is difficult to judge due to faintness, artifacts or other conditions. Here the measurements may be more uncertain than typical. The delta-magnitude is the most difficult measurement.
\end{table*}

From the speckle observations, we get magnitude differences in the B-band (centered at 562 nm) and the R-band (centered at 832 nm). As $\Delta G$ was one of the parameters used for the target selection in Paper I, we investigated the consistency between the Gaia photometry and the speckle photometry. To do this, we started with the relation between absolute Gaia $G$ magnitude and temperature from \cite{pecaut2013}. We then used the PHOENIX atmospheric models \citep{husser2013} corresponding to those temperatures to get the synthetic speckle $B$ and $R$ magnitudes relative to these $G$ magnitudes. With this synthetic photometry, we are able to get a relationship between the $\Delta B$ and $\Delta R$, and the modeled $\Delta G$ values. We examined the difference in the observed and predicted $\Delta B$ and $\Delta R$, where these values are predicted from our observed $\Delta G$ values, as a function of the separations found from the speckle observations. Here we find for $\rho < 0.6"$, that there is a larger scatter between the predicted and observed values, while for $\rho > 0.6"$ there is a constant difference between the predicted and observed values. For the short separations, this indicates that the flux in the Gaia $G$ band may be overestimated due to either poor PSF fitting or mis-identifications of the components at select epochs \citep{holl2023}. We examined what influence this effect would have on the target selection for Paper I, and found no significant change in the overall binary catalog. For the larger separations, we found the difference between the predicted and observed magnitudes in the speckle B-band to be $\sim1$ mag, indicating that the relative photometry may be unreliable. Because of these issues, we will not be using the speckle relative photometry in the subsequent analysis.

\section{Results}

In the current study, we only have two epochs of observation for each of our likely binaries; one from Gaia and the other from our speckle observations (Table \ref{tab:speckle_results}). With such little data, it is difficult to determine if the apparent motion between the epochs is due to orbital motion or relative motion due to chance alignment. In the following sections, we present methods to identify likely binaries based on their motion between these two epochs and analyze how this selection would change as a function of the errors on the positions.

\subsection{Characteristic Motion of Binaries}\label{sec:orb_mot}

One way to identify likely on-sky motion of a binary where little of its orbit has been observed was presented in \cite{Tokovinin2016}. In this method, the on-sky motion of the binary is compared to the motion of a face-on, circular orbit with a semi-major axis equal to the projected physical separation of the system, e.g., $a = s = \rho / p$. With this assumption, the characteristic speed of such an orbit can be described as:
\begin{equation}\label{eq:mu_star}
	\mu^\star = 2 \pi \rho^{-1/2} p^{3/2} M^{1/2}
\end{equation}
In the above, $\rho$ is the separation in arcseconds, $p$ is the parallax in arcseconds and $M$ is the sum of the mass of the system in Solar masses.

\cite{Tokovinin2016} then compared this characteristic motion to the observed motion of the system. Here the motion of the system was approximated as a linear function over time:
\begin{equation}
	\theta(t) \approx \theta_0 + \dot{\theta} (t - t_0)
\end{equation}
\begin{equation}
	\rho(t) \approx \rho_0 + \dot{\rho} (t - t_0)
\end{equation}
In the above, $t_0$ is the average time of observations. By solving for the above unknowns with our two data points for each system, the motion of the system is given by:
\begin{equation}
	\mu_t = \rho_0 \dot{\theta}
\end{equation}
\begin{equation}
	\mu_r = \dot{\rho}
\end{equation}
\begin{equation}
	\mu = \sqrt{\mu_t^2 + \mu_r^2}
\end{equation}
With this, the metric used for measuring the characteristic motion of a binary is $\mu^\prime = \mu / \mu^\star$. In \cite{Tokovinin2016}, $\mu^\prime$ is strictly defined for the \textit{instantaneous} separation and relative speed. With this definition, it can be shown analytically that for bound binaries $\mu^\prime < \sqrt{2}$, which \cite{Tokovinin2016} demonstrated through simulations to be true.

With the above in mind, we now calculate the characteristic motion, $\mu^\prime$, of the binaries with speckle observations. For all binaries, the separation and position angle were averaged between the red and blue band measurements to calculate the characteristic motion. Additionally, the position angle quadrant was chosen to match that of the Gaia detection. Also, to estimate the total mass of the system for eq. \ref{eq:mu_star}, we use the absolute magnitude of each source to estimate the mass based on the stellar mass-absolute magnitude relation from \cite{pecaut2013}. The $\mu^\prime$ values for our candidate binaries are listed in Table \ref{tab:mup_values}. Ignoring the speckle detection of an apparent third component in the field, we find that three candidate binaries have $\mu^\prime > \sqrt{2}$: Binary 1, Binary 9 and Binary 10.

\begin{table}
	\centering
	\caption{Characteristic motion, $\mu^\prime$, of the binaries with speckle observations. For all binaries, the separation and position angle were averaged between the red and blue band measurements to calculate the characteristic motion. Additionally, the position angle quadrant was chosen to match that of the Gaia detection. For reference, the last two columns show the projected physical separation, $s$, and the expected period of the system (assuming a circular orbit) based on the Gaia separation and parallax. For calculating the period, the mass of each component is estimated from the absolute magnitude mass relations from \cite{pecaut2013}.}
	\label{tab:mup_values}
	\begin{tabular}{lccc}
			\hline
			Binary Number & $\mu^\prime$ & $s$ & $P^\star$ \\
			&        &    [AU]  &     [years]      \\
			\hline
			1 & 3.0348 & 28.478 & 156.94\\
			2 & 0.7452 & 26.752 & 183.87\\
			3 & 0.3962 & 28.415 & 244.78\\
			4 & 0.4693 & 16.077 & 100.50\\
			4b$^\dagger$ & 6.4917 & \nodata & \nodata\\
			5 & 0.3592 & 19.374 & 146.02\\
			6 & 0.6404 & 12.754 & 64.44\\
			7 & 0.6795 & 14.388 & 65.84\\
			8 & 0.3736 & 10.476 & 45.66\\
			9 & 2.4954 & 14.055 & 69.19\\
			10 & 2.3169 & 17.225 & 82.79\\
			11 & 0.5576 & 15.940 & 78.94\\
			12 & 0.7709 & 10.930 & 42.90\\
			13 & 0.5174 & 17.754 & 109.38\\
			14 & 0.5225 & 13.880 & 59.79\\
			15 & 0.8210 & 17.014 & 79.44\\
			16 & 0.3077 & 16.894 & 86.12\\
			16b$^\dagger$ & 10.0345 & \nodata & \nodata\\
			\hline
	\end{tabular}
	\\
	$^\dagger$Speckle detection of an apparent third component in the field that was not detected in the original Gaia-2MASS binary analysis.
\end{table}

The large $\mu^\prime$ values for these three candidate binaries could be occurring for a couple of reasons. In the context of this study, $\mu^\prime$ is defined using the position differences rather than the instantaneous speed from \cite{Tokovinin2016}, meaning the same analytical limit may not hold. Additionally, if the errors are large in the Gaia and/or speckle positions, this could manifest in apparent motions that seem too fast for bound pairs. For the Gaia data, it is possible that mis-identifications could be occurring, i.e. swapping of the components between consecutive scans. This swapping would then manifest in both errors in the astrometry, which would mostly effect the position angle of the system, and decrease the magnitude difference of the pair \citep{holl2023}. For the speckle data, as is evident from the astrometric calibrations (Figure \ref{fig:asrtro_calib}), there could be large errors in both the separation and position angle depending on the conditions of a particular observing campaign. Because of these differences, we will redo the simulations from \cite{Tokovinin2016} to assess the true range of $\mu^\prime$  for binaries in the context of our observations. Effectively, these simulations will provide us the likelihood of a system being a binary given $\mu^\prime$; $P(\mu^\prime|B)$.

To do this, we generate 10,000 random orbits in a similar manner as \cite{Tokovinin2016}. As the period and semi-major axis do not effect the shape of the orbit, they are left as dimensionless quantities. The rest of the parameters are drawn from uniform distributions, where the longitude of periastron is over the range of $[0^\circ, 360^\circ]$, the eccentricity over $[0, 0.95]$, the mean anomaly over $[0^\circ, 360^\circ]$, the longitude of the ascending node over $[0^\circ, 360^\circ]$, and the inclination over $[0^\circ, 90^\circ]$ (where it is made to be uniform in $cos(i)$). Additionally, the time at which we are observing the binary is randomly selected from any position within the orbit. With these 10,000 random orbits, we then find the $\mu^\prime$ distribution for each binary for a range of assumed errors on the measurements. For each binary, we use the expected period of the system, assuming a circular orbit (Table \ref{tab:mup_values}), to find the time between epochs as a fraction of the total period. For the errors, we only apply them to the simulated observation at the speckle epoch, and assume they are random and Gaussian. This means they will act as the \textit{total} error on the positions, so they account for the errors in the Gaia and speckle data. For the simulated positions and position angles, we will probe errors of 0\%, 1\%, 5\%, 10\% and 20\%. Here, these errors are defined as the relative measurement errors on the separation and position angle, i.e. $\sigma_\rho / \rho$ and $\sigma_\theta / (2 \pi)$. With these assumptions and working in these dimensionless values for the period and semi-major axis, the characteristic motion is then given by $\mu^\prime = \mu \rho^{1/2} / (2 \pi)$, where $\mu$ and $\rho$ are the ``observed" motion and angular separation of the simulated binary, respectively. For getting the relative positions for each random orbit, we used the \texttt{twobody} Python package\footnote{\url{https://github.com/adrn/TwoBody}}.

\begin{figure*}[!t]
	\centering
	\includegraphics[width=\textwidth]{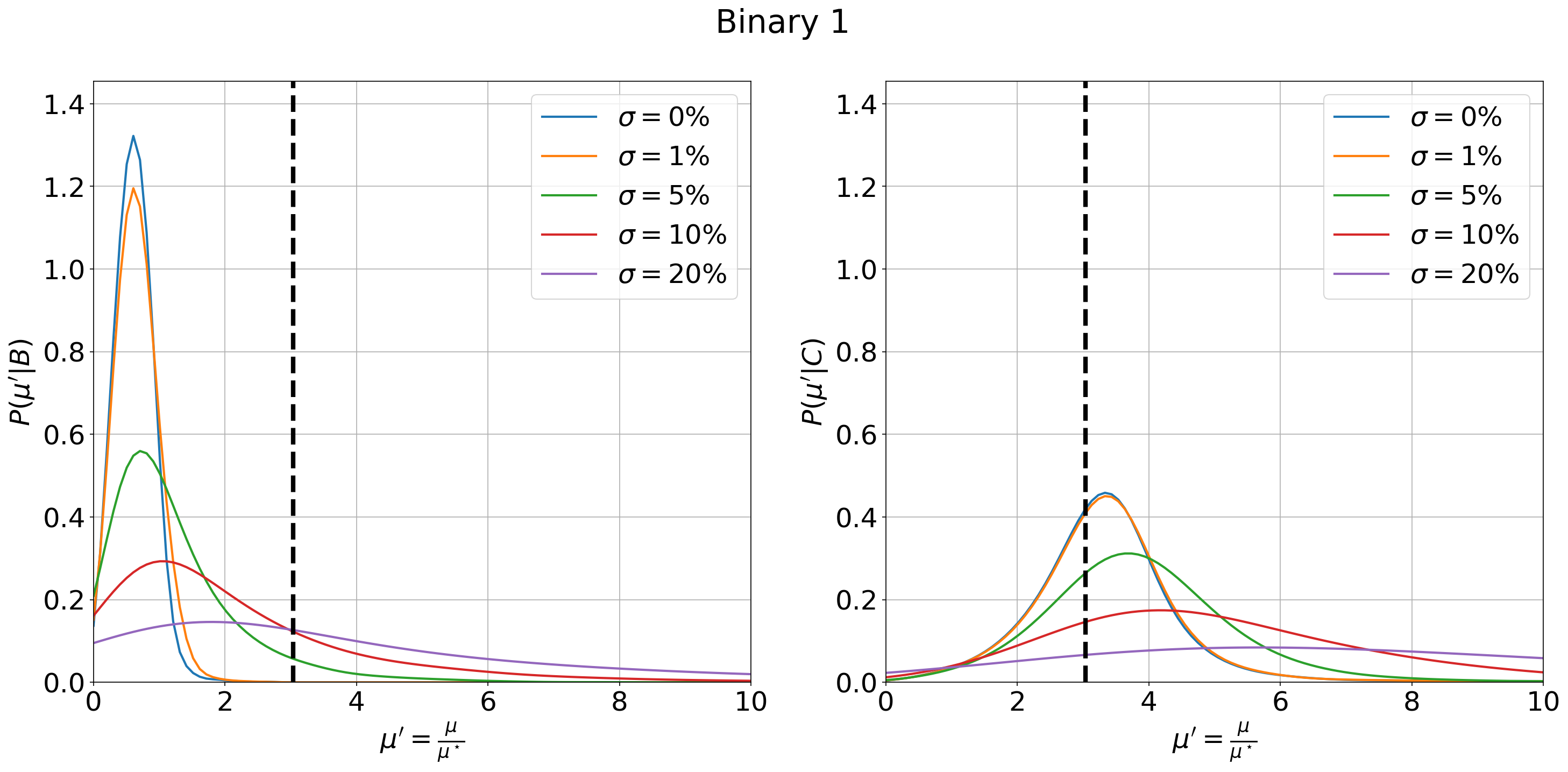}
	\caption{\textbf{Left Panel:} The likelihood of a system being a binary for a value of $\mu^\prime$, $P(\mu^\prime|B)$, for Binary 1. This likelihood is simulated from 10,000 random orbits drawn from uniform distributions, where the longitude of periastron is over the range of $[0^\circ, 360^\circ]$, the eccentricity over $[0, 0.95]$, the mean anomaly over $[0^\circ, 360^\circ]$, the longitude of the ascending node over $[0^\circ, 360^\circ]$, and the inclination over $[0^\circ, 90^\circ]$ where is is made to be uniform in $cos(i)$. The likelihood distributions shown are smoothed using a Kernel Density Estimator with a Gaussian kernel. The different colored solid lines show the likelihood for various assumed percent errors applied to the simulated speckle separations and position angles, i.e. $\sigma_\rho / \rho$ and $\sigma_\theta / (2 \pi)$. The dashed black line is the value of $\mu^\prime$ for Binary 1 from Table \ref{tab:mup_values}. \textbf{Right Panel:} The likelihood of a system being a chance alignment for a value of $\mu^\prime$, $P(\mu^\prime|C)$, for Binary 1. To generate these likelihood distributions, 10,000 field stars from a $5^\circ$ region around the primary source are selected. By fixing the proper motion of the primary and using the proper motions of these field stars as that of the secondary, the relative motion between the Gaia and speckle epochs can be found.  The likelihood distributions shown are smoothed using a Kernel Density Estimator with a Gaussian kernel. The different colored solid lines show the likelihood for various assumed percent errors applied to the simulated speckle separations and position angles, i.e. $\sigma_\rho / \rho$ and $\sigma_\theta / (2 \pi)$. The dashed black line is the value of $\mu^\prime$ for Binary 1 from Table \ref{tab:mup_values}.}
	\label{fig:mu_prime_orbs}
\end{figure*}

The result of this simulation procedure for Binary 1 in shown in the left panel of Figure \ref{fig:mu_prime_orbs}. Here the likelihood distributions have been estimated using Kernel Density Estimator with a Gaussian kernel, as implemented in \texttt{scikit-learn} \citep{scikit-learn}. We estimate the bandwidth of the kernel using the method from \cite{scott1992} and multiply this value by three to ensure smoothness of the likelihood distribution at regions of low sampling. For Binary 1, it is clear that if the errors on the speckle measurements are very small (1\%), then it is very unlikely to have such high value of $\mu^\prime$ (black dashed line in Figure \ref{fig:mu_prime_orbs}). As we increase the errors on the speckle positions though, the distributions begins to shift to larger values of $\mu^\prime$ and the likelihood of having a large $\mu^\prime$ value increases. This is a trend that is consistent for likelihood functions simulated for each candidate binary. This demonstrates that large errors could be an explanation for the large $\mu^\prime$ values found for three of our candidate binaries. To know if this is the sole reason though, we must consider the alternative hypothesis; that the large motion is due to a chance alignment. We will consider this hypothesis in the next section.

\subsection{Probable Location of Background Star}\label{sec:prop_back}

One aspect of the previous section that is not considered is the proper motion of the primary star for each candidate binary. By combining this with the typical proper motion of field stars, one can determine the probable location of the secondary at the speckle epoch to see if this motion is consistent with a chance alignment. Doing this analysis will help determine the likelihood of a system being a chance alignment given $\mu^\prime$; $P(\mu^\prime|C)$.

To test this, we query Gaia DR3 in a $5^\circ$ region around the primary target. As faint, distant sources will have a distinct proper motion distribution compared to the brighter sources more consistent with the true secondary stars, we only select objects in this $5^\circ$ region where the $G$ magnitude is less than that of the secondary. Many of these sources will have proper motions near zero, but with a diffuse halo of higher proper motion objects that, while still unrelated to the primary, are closer in distance from the Sun or have higher space velocities. To then get the expected distribution of the location of the secondary if it was a background field star at the epoch of the speckle observations, we do the following.

We randomly select 10,000 field stars from the $5^\circ$ region around the primary source and assume these proper motions to be the proper motion of the secondary for each iteration. In each of these iterations, we also attempt to take into account the errors on the $(\alpha, \delta)$ of the primary and secondary, and the proper motion of the primary by randomly varying these locations and proper motion assuming the errors are Gaussian. With these locations and proper motions, we then recalculate the location of the primary and secondary at the epoch of the speckle observations. Here we then additionally apply relative, random errors of 0\%, 1\%, 5\%, 10\% and 20\% to the separation and position angle at the epoch of speckle of observation, similar to the simulations in Section \ref{sec:orb_mot}. By repeating this for 10,000 iterations, we build up a distribution of the likely relative position of the secondary assuming it is unrelated to the primary source. With these positions, we then calculate the corresponding $\mu^\prime$ value using the same assumptions as were used for our candidate binaries. It is important to note here that this means we will swap the quadrant of the position at the epoch of speckle observation to match the Gaia one.

We repeat the above procedure for each candidate binary to construct our likelihood distributions for being a chance alignment. As in Section \ref{sec:orb_mot}, we approximate the likelihoods using a Kernel Density Estimator with a Gaussian kernel. An example of such a distribution for Binary 1 is shown in the right panel of Figure \ref{fig:mu_prime_orbs}. Similar to the distributions of $P(\mu^\prime|B)$ (left panel, Figure \ref{fig:mu_prime_orbs}), as the percent error on the simulated speckle position increases, the distribution generally skews to larger values of $\mu^\prime$ and has a much larger variance. For this candidate binary though, it is clear that if the errors are small on the speckle position, then there is a high likelihood that the observed motion is consistent with a chance alignment. This becomes less likely as the errors increase though.

With these distributions we can determine the probability of a candidate binary being a physical system based on its observed $\mu^\prime$. This is calculated in a Bayesian sense as:
\begin{equation}
	P(B|\mu^\prime) = \frac{P(\mu^\prime|B) P(B)}{P(\mu^\prime|B) P(B) + P(\mu^\prime|C) P(C)}
\end{equation}
In the above, the likelihoods for each binary come from the process described in Section \ref{sec:orb_mot} and \ref{sec:prop_back}, and the priors will come from the expected binary rate from Paper I; $P(B) = 0.996$ and $P(C) = 1 - P(B)$. Combining this with the $\mu^\prime$ values for each candidate binary (Table \ref{tab:mup_values}), we get the posterior probability of the candidate binary being a physical system for an assumed total percent error on the measurements in Table \ref{tab:prob_binary}. From this analysis, only Binary 1 has a possibility of being chance alignment if we require a high probability threshold ($>95\%$) and assume the errors on the measurements are small ($\sim1\%$); all other candidate binaries have a high likelihood of being physical systems regardless of the errors on the measurements.

\begin{table*}
	\centering
	\caption{The Bayesian probability of a candidate binary being a physical system,  $P(B|\mu^\prime)$, given an assumed percent error on the simulated position at the time of the speckle observation.}
	\label{tab:prob_binary}
	\begin{tabular}{p{5pt}c|cccccccccccccccc}
			\multicolumn{2}{c}{} & \multicolumn{16}{c}{Binary Number} \\
			& & 1 & 2 & 3 & 4 & 5 & 6 & 7 & 8 & 9 & 10 & 11 & 12 & 13 & 14 & 15 & 16  \\
			\hline
			\multirow{8}{*}{\rotatebox[origin=c]{90}{Percent Error}}
			& 0\%   & 0.030 & 1.000 & 1.000 & 1.000 & 1.000 & 1.000 & 1.000 & 1.000 & 0.999 & 0.963 & 1.000 & 1.000 & 1.000 & 1.000 & 1.000 & 1.000 \\
			& 1\%   & 0.027 & 1.000 & 1.000 & 1.000 & 1.000 & 1.000 & 1.000 & 1.000 & 1.000 & 0.963 & 1.000 & 1.000 & 1.000 & 1.000 & 1.000 & 1.000 \\
			& 5\%   & 0.982 & 1.000 & 1.000 & 1.000 & 1.000 & 1.000 & 1.000 & 1.000 & 1.000 & 0.995 & 1.000 & 0.999 & 0.999 & 0.999 & 1.000 & 1.000 \\
			& 10\% & 0.995 & 1.000 & 1.000 & 1.000 & 1.000 & 1.000 & 1.000 & 1.000 & 1.000 & 0.998 & 1.000 & 0.996 & 0.998 & 0.999 & 1.000 & 1.000 \\
			& 20\% & 0.998 & 1.000 & 0.999 & 0.999 & 0.999 & 1.000 & 1.000 & 0.999 & 1.000 & 0.998 & 1.000 & 0.996 & 0.998 & 0.998 & 1.000 & 1.000 \\
			\hline
	\end{tabular}
\end{table*}

\subsection{Third Components Detected in the Speckle Field}\label{sec:trips}

\begin{figure*}[!t]
	\centering
\gridline{\fig{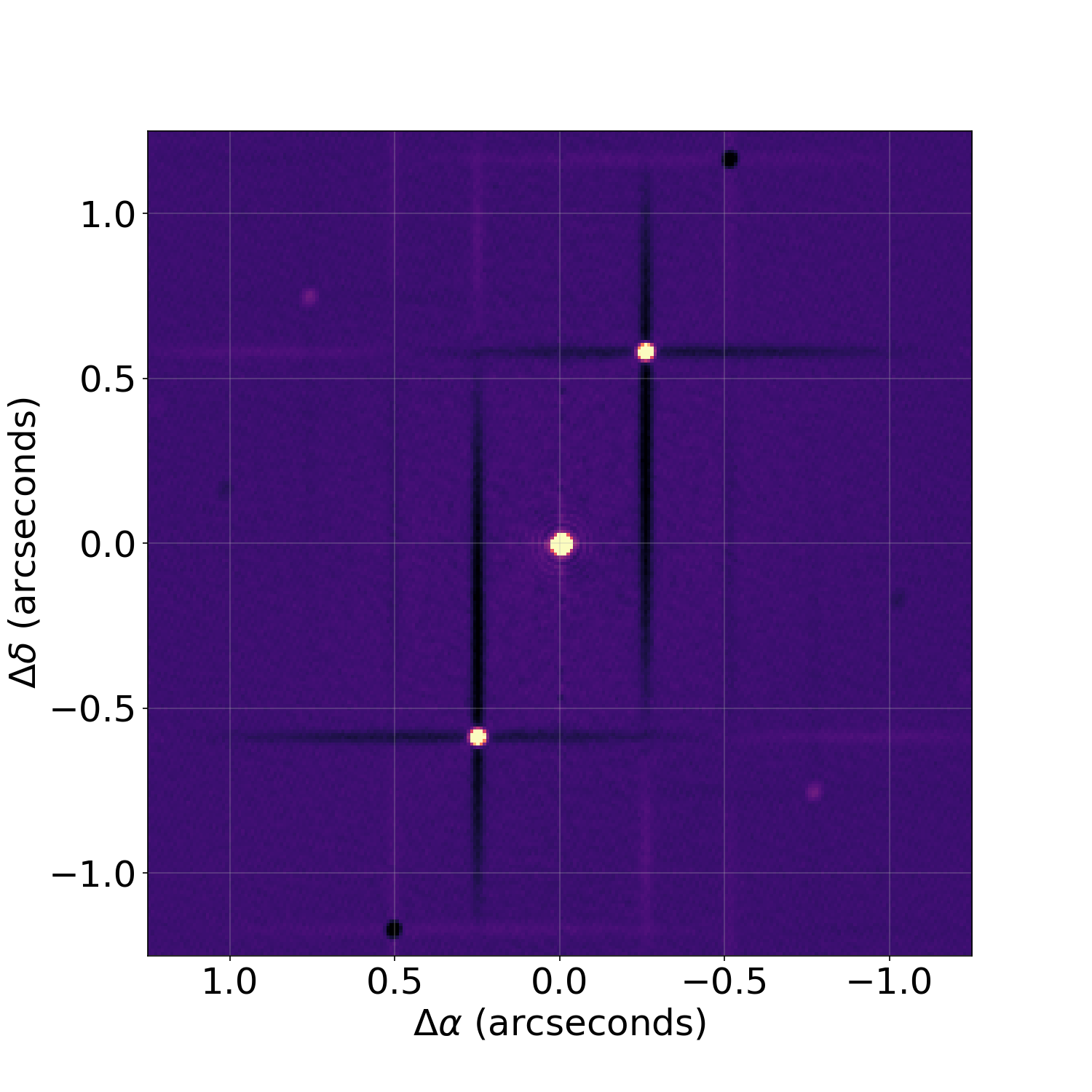}{0.45\textwidth}{Binary 4b}
	\fig{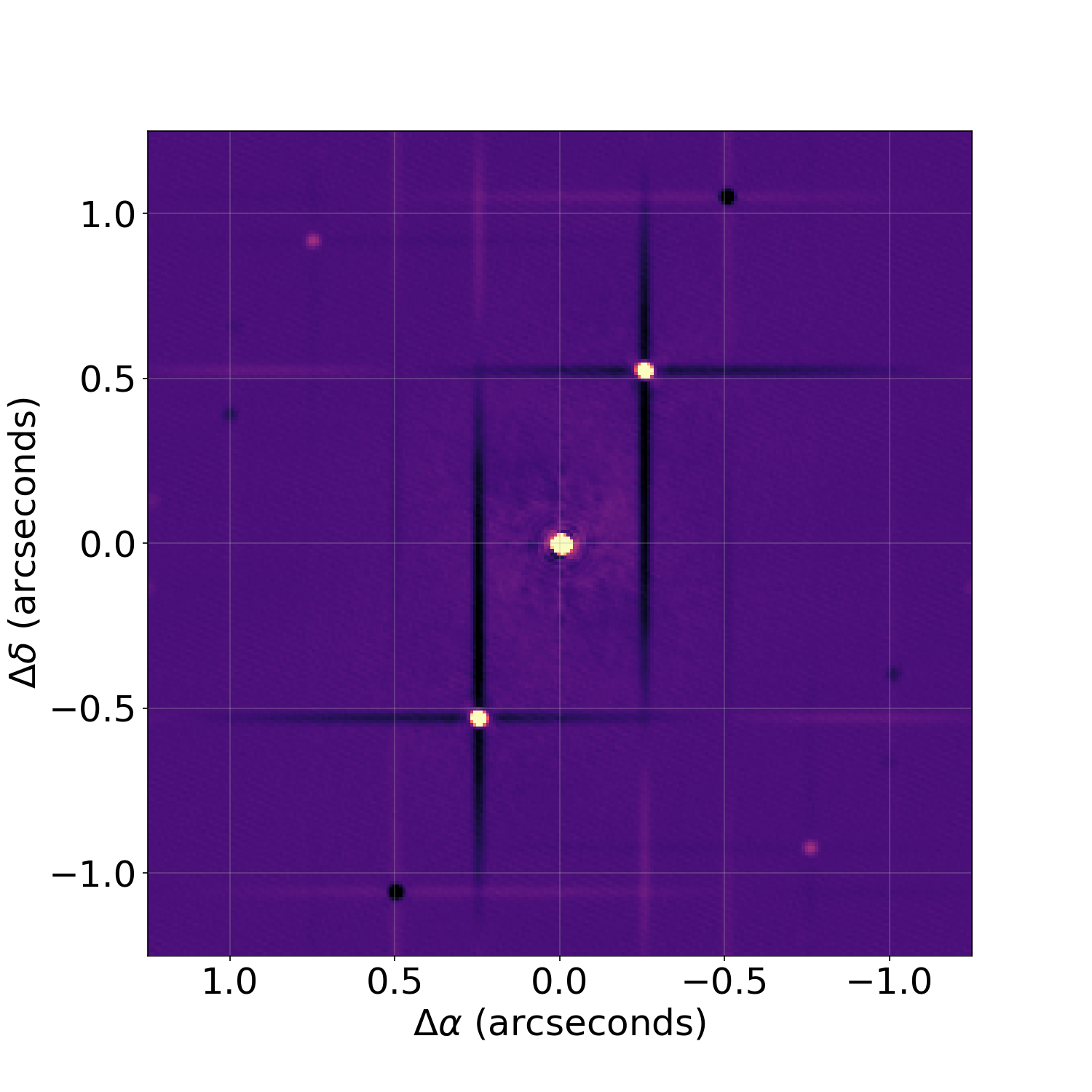}{0.45\textwidth}{Binary 16b}}
	
	\caption{Here we show the reconstructed images in the 832 nm band from the speckle pipeline for each of the binaries with a third component detected in the speckle field.}
	\label{fig:speckle_results_trips}
\end{figure*}

For two of the candidate binaries (Binary 4 and Binary 16), the speckle analysis finds a third component within the FOV of the observations. The reconstructed images for these two systems are shown in Figure \ref{fig:speckle_results_trips}, where the faint third components can be seen at larger separations than the secondary. We searched the Gaia archive around the primary both at epoch $=2016$ and at the epoch of observation, and did not find any other sources within the FOV of the observations, so if these are field stars they do not show up in Gaia. Additionally, we searched outside of the speckle field to see if there were any nearby bright sources, as these could show up as aliased peaks in the speckle FOV. From this, we do not find any other bright stars with 10” of either of these systems, though we do find one $G=20$ mag source within 10" of Binary 16. This indicates that these are true detections and not aliased peaks from other sources. We do note that these sources are found to be much fainter than the primary, so it is likely they are too faint to be detected in Gaia. Also, these sources are so faint that depending on observing conditions, they may not be able to be detected with the speckle instruments of Gemini as they lie at the detection limit. Indeed, for Binary 4, which we have two speckle epochs of observation, the third component was only detected for one of these epochs. Additionally, due to the orientation of the systems they both seem more consistent with a chance alignment with a field star, as we normally expect triples to be hierarchical. Overall, with the available data we consider it most likely that these third components are chance alignments and are not associated with the binary systems considered.

\subsection{Multiple Epochs of Observation}\label{sec:multi}

We have one system (Binary 4) where we were able to get an additional epoch of observation, approximately 5 months after the initial one. In the red band (as the reduction was not successful for the blue band at the second epoch), we observed a change in position angle between the two speckle epochs of $\sim0.055^\circ$, where this is based on the average position angle between each band for each epoch. We do note that this change is an order of magnitude smaller than the astrometric precision for the observing runs (Table \ref{tab:astro_calib}), so this should be considered in the following.

If we assume the orbit is circular and face on, we can estimate the amount of apparent orbital motion expected for the time difference between these two epochs. Here we assume we can convert the absolute G magnitude of each of the components to mass using the relation from \citep{pecaut2013} and when combining this with the separation from the speckle results (Table \ref{tab:speckle_results}), we get a period of $\approx 74.97$ years. This means that for this time difference, we would expect to have completed 0.52\% of the orbit. Based on the difference in position angle however, we find that (assuming a face on circular orbit) the system has only completed 0.015\% of the orbit. Again, due to the astrometric precision though the difference in position angle could be much greater. With an average precision for these observing runs of $\sim0.6^\circ$, this means that at the $1\sigma$ level the system could have completed $0.25\%$ of the orbit. This is still smaller than what is estimated, but it does not rule out orbital motion and may simply mean the system is not face on. Longer baseline follow-up observations are thus needed to continue to monitor the orbital motion of this system, and all systems here that seem to not be associated with the motion of a background field star.

\section{Discussion}\label{sec:discuss}

From our speckle observations, we have been able to detect secondary sources in all images for the binary candidate targets. In order to assess if the apparent motion between the Gaia data and the speckle observations was consistent with a physical companion, we calculate the characteristic motion, $\mu^\prime$ for all systems. To determine if this motion is consistent with a physical system, in Section \ref{sec:orb_mot} we created a series of likelihood functions generated from random orbits where the simulated separation and position angle at the speckle epoch were given some level of random error. To test the alternative hypothesis, in Section \ref{sec:prop_back} we created similar likelihood functions for what this same characteristic motion would look like for chance alignments. These likelihood distributions allowed us to find the probability that a candidate binary is a physical system given its observed $\mu^\prime$ (Table \ref{tab:prob_binary}). Here we discuss the implication of these results.

For some assumed total error on the positions and a probability cut on $P(B|\mu^\prime)$, we could validate the method used in Paper I for selecting binary candidates from Gaia DR3 and estimate  the likely fraction of true binaries from the proposed list of candidates. Here, the sample proportion of detected binaries would be:
\begin{equation}
	\hat{p} = \frac{x}{n}
\end{equation}
The standard error on this measurement is then:
\begin{equation}
	SE =z \sqrt{\frac{\hat{p} (1 - \hat{p})}{n}}
\end{equation}
Where $z=1.96$ for a 95\% confidence interval. If we use a threshold of $P(B|\mu^\prime) > 95\%$, this implies that if the errors are small ($\sim1\%$), then 15 out of the 16 candidate binaries are physical systems and our detection rate is $93.75 \pm 11.86 \%$. Our original predicted fraction of binaries within the population for the ``clean" sample, $99.6\%$, falls withing this range, but due to the small sample size this would not significantly confirm this predicted value. However, even in this outcome the high fraction of binaries recovered in our small study strongly suggests that the majority of candidates binaries presented in Paper I are in fact physical systems, and thus warrant further investigation.

It seems unlikely that the errors are this small given the data though. If we assume all candidate binaries are physical systems, we can determine the likely total error on the measurements assuming some average value across all observations. Here we slightly redo the likelihood calculations from Section \ref{sec:orb_mot}, where for the 10,000 random orbits we select times between epochs as a fraction of the total period that follow the same distribution as the observed sample. Additionally, we calculate $\mu^\prime$ values for these 10,000 random orbits assuming total errors between 0\% and 20\%, and use them to find the likelihood distribution at each of these total errors. Using the observed $\mu^\prime$ values (Table \ref{tab:mup_values}) and these distributions, we find that the marginal likelihood is maximized for a distribution with a total, relative error on the separations and position angles (i.e. $\sigma_\rho / \rho$ and $\sigma_\theta / (2 \pi)$) of $\sim4.3\%$. At this level of error, indeed we would conclude that 16 of the 16 binaries are physical systems given a threshold of $P(B|\mu^\prime) > 95\%$. If Binary 1 is indeed a chance alignment though and we do not include it when calculating the maximum likelihood, a distribution with a total error of $\sim3\%$ best describes the sample.

This would imply that the large apparent motions for some of the candidate binaries are simply due to erroneous positions from Gaia and/or the Gemini speckle data. These errors would have two likely sources. For the Gaia positions, due the short separations of the systems and their small magnitude differences, depending on the scan angle of the telescope relative to the position angle of the binary, either ``swaps" (mis-identification of the sources) or blending of the two objects can occur \citep{holl2023}. This can lead to erroneous positions in the Gaia catalog, where the largest errors would most likely be in the position angle of the system. Such errors make sense for our candidate binaries as they all have high RUWE values, which indicates poor astrometric fits and thus less certain positions.

For the speckle data, astrometric calibration is key to achieving results with a high level of precision. Issues with, say, the pixel scale are multiplicative in nature. As a result, even a 1 mas error in the plate scale can lead to $\sim 10\%$ error, e.g., a 10 mas error for a 100 mas separation binary. Here we find that the pixel scale variance per observing run is an order of magnitude less than this, so we do not expect the errors on the speckle data alone to be this high. When combined with the errors in the Gaia positions though, it seems conceivable that the average total error could be as high as $\sim4.3\%$.

This is a simplistic view of the errors on the measurements though. We do not expect that all measurements are drawn from the same error distribution. Indeed, the errors are going to be a function of the separation of the system, magnitude difference between components, the position angle relative to the Gaia scan angles and the observing conditions for the speckle observation. This means that the total error on some measurements may be less and some greater. An average error as high as $\sim4.3\%$ does have implications for any orbital solutions derived from observations with Gaia and/or Gemini speckle instruments though. Even at this moderate level, such errors could have significant effects on dynamical masses derived from such orbital solutions. So, in the absence of robust error estimation on the positions of close separation binaries from Gaia and/or Gemini speckle observations, we strongly suggest that this average relative error of $\sim4.3\%$ should be considered in all subsequent analyses with such data.

\section{Conclusions}

In this study, we observed 16 candidate binaries with small physical separations ($s<30$ AU) from the catalog of high probability binaries presented in the first paper of this series \citep[Paper I;][]{medan2023}. The main goals of this work were to (1) confirm the systems as physical binaries and (2) validate the method from Paper I by assessing the contamination rate from background field stars in the current sample.

We observed the 16 candidate binaries using the ‘Alopeke and Zorro speckle cameras on the Gemini $8.1$ m North and South telescopes, respectively, from February 2022 to January 2023. In all observations, we detected a secondary component around the primary source. In addition to these detections, some of our speckle observations also revealed a third source in the FOV for two of the candidate binaries (Binary 4 and 16). From querying the Gaia catalog, we do not find any additional sources in the field that could explain these observations, though we do find that these third sources are much fainter than the primary so it is likely they are too faint to be in Gaia. With the current data though, it seems most likely that these are just chance alignments to the candidate binaries.

Also in all observations, there appears to be motion between the Gaia epoch and the observation date with the speckle camera. To determine if this motion is more consistent with orbital motion in a physical binary, or whether it is more consistent with relative motion of a random field star, we implemented the following. First, we calculate the characteristic orbital motion statistic, $\mu^\prime$, from \cite{Tokovinin2016}. From this we find that three binaries have $\mu^\prime > \sqrt{2}$, which is the limit for bound systems. To assess the effect differing levels of error may have on our measurements, we simulate 10,000 random binaries. Here we add varying levels of random error and demonstrate that large measurement errors can result in $\mu^\prime$ values higher than expected for bound systems.

The alternative hypothesis is that the large motions are caused by chance alignments. To test this, we use the proper motion of stars within $5^\circ$ of the primary star to determine the probable location of the secondary if its motion was similar to that of a chance alignment with a typical field star. Similar to the above, we add varying levels of random error and find the likelihood distribution of being a chance alignment for each binary. By combining this with the likelihood of a system being a binary for some value of $\mu^\prime$, we find that 15 of the 16 binaries are physical systems regardless of the level of error. This would imply that $93.75 \pm 11.86 \%$ of the candidates in our sample are likely physical (true) binaries, at a confidence level of 95\%. For this to be true, than the total, relative error on the separation and position angle measurements (i.e. $\sigma_\rho / \rho$ and $\sigma_\theta / (2 \pi)$) would have to be $<3\%$. We discuss how this is unlikely given our knowledge of the Gaia and Gemini speckle data reductions. Indeed, our findings accommodate all 16 binaries as physical system if the total average measurement error is $\sim4.3\%$.  Based on our current knowledge of the Gaia and speckle pipelines, such level of total average error seems conceivable.

Finally, we also collected one additional epoch of observation for Binary 4 approximately 5 months after the initial speckle observation. Here we observe some motion between these two epochs that (assuming a face on circular orbit) indicates the secondary has completed 0.015\% of the orbit. This is much smaller than what is estimated for a circular orbit with this physical separation, but it does not rule out orbital motion and may simply mean the system is not face on. 

This demonstrates that in this system, and all the binaries in our sample, longer baseline follow-up observations over the next decades will be needed to completely monitor their orbits. Such monitoring is crucial, as it allows for mass determinations of each component of the system. These masses contribute to the growing statistics of low-mass binaries in the Solar Neighborhood and aid in the bettering of binary formation and evolution theory. One important consideration for such monitoring though is the level of error on the measurements. From this work we find that, on average, the total relative measurement error on the binary separations and position angles is $\sim4.3\%$.  This demonstrates that beyond using this data to assess the likelihood of a system being a true binary vs. a chance alignment, this quantitative assessment of the true average measurement error allows for more robust error estimates of mass determinations from short separation binaries with Gaia and/or Gemini speckle data.

\section*{Acknowledgments}

Dr.~Medan gratefully acknowledges support from a Georgia State University Second Century Initiative (2CI) Fellowship and a Vanderbilt Initiative in Data-Intensive Astrophysics (VIDA) Fellowship.

We greatly thank the anonymous referee for their comments and suggestions, which greatly improved our manuscript.

Observations in the paper from programs GN-2022A-Q-313, GS-2022A-Q-316 and GS-2022B-Q-313 made use of the High-Resolution Imaging instrument(s) ‘Alopeke (and/or Zorro). ‘Alopeke (and/or Zorro) was funded by the NASA Exoplanet Exploration Program and built at the NASA Ames Research Center by Steve B. Howell, Nic Scott, Elliott P. Horch, and Emmett Quigley. ‘Alopeke (and/or Zorro) was mounted on the Gemini North (and/or South) telescope of the international Gemini Observatory, a program of NSF’s NOIRLab, which is managed by the Association of Universities for Research in Astronomy (AURA) under a cooperative agreement with the National Science Foundation on behalf of the Gemini Observatory partnership: the National Science Foundation (United States), National Research Council (Canada), Agencia Nacional de Investigaci\'{o}n y Desarrollo (Chile), Ministerio de Ciencia, Tecnolog\'{i}a e Innovaci\'{o}n (Argentina), Minist\'{e}rio da Ci\^{e}ncia, Tecnologia, Inova\c{c}\~{o}es e Comunica\c{c}\~{o}es (Brazil), and Korea Astronomy and Space Science Institute (Republic of Korea).

This work has made use of data from the European Space Agency (ESA) mission
{\it Gaia} (\url{https://www.cosmos.esa.int/gaia}), processed by the {\it Gaia}
Data Processing and Analysis Consortium (DPAC,
\url{https://www.cosmos.esa.int/web/gaia/dpac/consortium}). Funding for the DPAC
has been provided by national institutions, in particular the institutions
participating in the {\it Gaia} Multilateral Agreement.

\bibliographystyle{aasjournal}
\bibliography{gaiadr3_binaries_speckle}

\end{document}